%% file: main.tex
\documentclass[conference,a4paper]{IEEEtran}
\usepackage{amsmath,amsfonts,amssymb,amsthm, bm}
\usepackage{graphicx} % Required for inserting images
\usepackage{tikz}
\usetikzlibrary{external}
\usepackage{pgfplots} 
\usepackage{pgfgantt}
\usepackage{pdflscape}
\pgfplotsset{compat=newest} 
\pgfplotsset{plot coordinates/math parser=false}
\usepackage{subcaption}
\usepackage{acronym}
\acrodef{OFO}{Online Feedback Optimization}
\acrodef{MPC}{Model Predictive Control}
\acrodef{OPF}{Optimal Power Flow}
\acrodef{MIQP}{mixed integer quadratic optimization problem}
\acrodef{TRL}{Technology Readiness Level}
\usepackage{cite}
\usepackage{adjustbox}
\newcommand{\bbR}{\mathbb{R}}

\newlength\fwidth
\newlength\fheight

\usepackage[hyperindex=true, %
  pdftitle={Deployment of an Online Feedback Optimization Controller for Reactive Power Flow Optimization in a Distribution Grid},%
  pdfauthor={Lukas Ortmann, Christian Rubin, Alessandro Scozzafava, Janick Lehmann, Saverio Bolognani, Florian Dörfler},%
  colorlinks=true,%
  pagebackref=false,%
  plainpages=false,%
  pdfpagelabels,%
  linkcolor=black,%
  citecolor=black,%
  filecolor=black,%
  urlcolor=black%
  ]{hyperref}

\title{Deployment of an Online Feedback Optimization Controller for Reactive Power Flow Optimization in a Distribution Grid}
\author{Lukas Ortmann, Christian Rubin, Alessandro Scozzafava, Janick Lehmann, Saverio Bolognani, Florian Dörfler
\thanks{Lukas Ortmann, Christian Rubin, Florian D\"{o}rfler, and Saverio Bolognani are with the Automatic Control Laboratory at ETH Zurich, Zurich, Switzerland. Email: $\{$ortmannl,dorfler,bsaverio$\}$@ethz.ch}
\thanks{Alessandro Scozzafava and Janick Lehmann are with AEW Energie AG (Swiss distribution grid operator), Aarau, Switzerland. Email: firstname.lastname@aew.ch}
\thanks{Research supported by the Swiss National Science Foundation under NCCR Automation, grant agreement 51NF40\_180545}
}
\makeatletter

\IEEEoverridecommandlockouts

\begin{document}
% \IEEEpubid{\makebox[\columnwidth]{979-8-3503-9678-2/23/\$31.00 \copyright 2023 IEEE \hfill} \hspace{\columnsep}\makebox[\columnwidth]{ }}

\maketitle
\begin{abstract}
Optimization is an essential part of power grid operation and lately, Online Optimization methods have gained traction. One such method is \ac{OFO} which uses measurements from the grid as feedback to iteratively change the control inputs until they converge to the solution of the optimization problem. Such algorithms have been applied to many power system problems and experimentally validated in lab setups. This paper implements an \ac{OFO} controller in a real distribution grid for 24/7 operation using off-the-shelf hardware and software. The proposed control strategy optimizes the reactive power flow at the substation while satisfying voltage constraints. As part of an existing coordination scheme between (sub)transmission grid operator (TSO) and distribution grid operator (DSO), this comes with a financial reward and simultaneously it virtually reinforces the grid by regulating the voltage on the feeder and therefore allowing higher levels of distributed generation/consumption. We present how a distribution grid is retrofitted such that we can use existing inverters, we analyze the controller's interaction with legacy infrastructure, and investigate its overall control behavior. Finally, we demonstrate the successful deployment of an OFO controller in an operational environment which corresponds to \ac{TRL}~7.
\end{abstract}

\section{Introduction}
The operation of power systems comprises many tasks that can be formulated as optimization problems
A famous example is Optimal Power Flow. Defining a control objective as an optimization problem is powerful, flexible, and versatile. Often, optimization problems even arise naturally e.g. when constraints, like voltage limits, need to be satisfied. It is therefore important to develop, deploy and evaluate different methods that can solve optimization problems under real operating conditions in a grid. More precisely, methods are needed that are fast and robust, and, especially in distribution grids, need to be able to work with little model information. If an exact model is available these optimization problems are solved offline on a computer by using an optimization algorithm and the model. The solution of the optimization is then deployed onto the grid, see Figure~\ref{fig:block_diagram_offline_opt}. However, solving these optimization problems offline can be computationally intense and they need to be robustified to be able to deal with model mismatch. Otherwise, a model mismatch could lead to a constraint violation. Unfortunately, such robustification prohibits to utilize the grid to its full capacity because some margin needs to be included to deal with a model mismatch. In distribution grids, no good system model might exist in the first place. 

To circumvent these problems, Online Optimization methods have been developed that take feedback into account, see Figure~\ref{fig:block_diagram_OFO} and consult~\cite{molzahn2017survey} for a detailed review. One such method is called \ac{OFO}. This method allows to steer a system to the solution of an optimization problem by taking decisions that are not based on an available model of the grid but on measurements collected in real-time. It is computationally light, robust to model mismatch, can utilize a grid to its full capacity, and needs very limited model information, see~\cite{hauswirth2021optimization} for a review paper. In simulations, it has been applied to a vast number of power system problems~\cite{olives2022model,olives2023holistic,tang2020measurement,ipach2022distributed,nowak2020measurement_PVSC,gan2016online,dall2016optimal,bernstein2019real} and it has also been experimentally tested with hardware-in-the-loop simulations~\cite{wang2020performance,padullaparti2021peak}. Experiments using a real power grid setup have also been done, however, those tests were either done in dedicated lab environments~\cite{ortmann2020experimental, ortmann2020fully,reyes2018experimental} or on microgrids~\cite{kroposki2020autonomous,kroposki2020good} using a specialized hardware and communication setup. In contrast, this paper presents the deployment of an \ac{OFO} controller in a real distribution grid for 24/7 operation utilizing existing hardware. The distribution grid we chose is operated by AEW Energie AG, is located in the north of Switzerland, and supplies 100.000 people.
The objective of the controller is twofold: On the one hand, it is tasked to optimize the reactive power flow at the substation, based on a TSO-DSO coordination scheme, that yields financial rewards for the distribution grids. On the other hand, it is used to regulate the voltage inside the distribution feeder. Such voltage support virtually reinforces the grid through automation and has the potential to mitigate or postpone grid reinforcements~\cite{Verteilnetzstudie}. The potential of such virtual grid reinforcement through coordinated reactive power sources was analyzed in~\cite{matt2023virtual} and the authors concluded that 9\% more active power can be conducted before voltage constraints limit the possible active power flow.

The paper documents an example of successful TSO-DSO coordination in the Swiss power system and provides a demonstration of the effectiveness of OFO for real-time optimization problems in the power grid.
Our contributions can be structured as follows: 1) we present how we retrofitted the distribution grid infrastructure both in terms of the communication and hardware setup, 2) we investigate the consequences of using off-the-shelf hardware and the interaction of a real distribution grid with an \ac{OFO} controller which serves as a robustness test of \ac{OFO}, 3) we give a tutorial on \ac{OFO}, including potential extensions of the controller, that will assist the power system community in using this new technology for other control and optimization problems.\\
Overall, with this deployment for 24/7 operation in a real distribution grid, \ac{OFO} has reached \ac{TRL}~7 ("system prototype demonstration in an operational environment")~\cite{ISO_TRL}.

\section{Reactive Power Prices in Switzerland}

The Swiss transmission grid operator, Swissgrid, is controlling its voltage with the help of generators and distribution grids that are connected to the transmission grid. Generators connected to the transmission grid have to participate in so-called active voltage support while subtransmission grid operators have to participate in so-called semi-active voltage support and they can opt-in for active voltage support. The basis for this voltage support scheme is that Swissgrid calculates a voltage reference for every bus in the transmission grid. This is done every 15~minutes through an Optimal Power Flow solver. All entities connected to the transmission grid are incentivized to adjust their reactive power flow such that it helps to drive the voltage at their connection point to the provided reference. The incentive scheme works as follows: Reactive power flows that are helping to drive the voltage to the reference are considered conform whereas reactive power flows that have the wrong sign and drive the voltage away from the reference are considered non-conform. In both active and semi-active voltage support, the generators and subtransmission grid operators are financially rewarded when they provide conform reactive power flows and they pay penalties when their reactive power flows are non-conform. The prices and penalties differ between active and semi-active voltage support. Furthermore, in semi-active voltage support, a tolerance band exists within which no reward nor penalty is billed. See~\cite{Spannungshaltung} for more information.\\
The subtransmission grid operators forward this pricing scheme to the distribution grid operators and charge or pay the distribution grids depending on the reactive power flow at the connection points between their subtransmission and the distribution grid. Hence, the distribution grid operators have a financial incentive to control their reactive power flows as well. This can be done with inverters and generators connected to the distribution grid as they can serve as reactive power sources. However, their reactive power injections have lower and upper limits ($q_{min}$ and $q_{max}$) due to the hardware limits of the inverters and generators. Furthermore, reactive power flows also affect the voltages in the grid and one needs to make sure that all voltages $v$ stay within their lower and upper limits ($v_{min}$ and $v_{max}$). Therefore, an optimization problem arises: How can reactive power injections $q$ be used to minimize the cost and maximize the reward from the subtransmission grid operator while satisfying the voltage and hardware limits. Mathematically speaking, we define the constraint optimization problem:
\begin{align}\label{eq:op_problem_AEW}
	       \begin{split}
	        	        \min_{q} \; &cost(q) - reward(q)\\
	        	        \text{s.t.}\quad &q_{min}\leq q \leq q_{max}\\
	        	        &v_{min} \leq v \leq v_{max}\\
                        &v=h(q,d)
	        	        % &u_i \in \mathbb{Z} \quad \forall \; \text{discrete inputs} \; i 
	        \end{split}
	        \end{align}
We will describe the relationship between $q$ and $v$ as $v=h(q,d)$ where $h(\cdot)$ represents the power flow equations and $d$ is a vector of all active and reactive injections in the grid.
The goal of our \ac{OFO} controller and its implementation in the distribution grid will be to iteratively change the reactive power injections $q$ until they converge to $q^\star$ that optimally solves the optimization problem~\eqref{eq:op_problem_AEW}.

\section{Online Feedback Optimization}
\begin{figure}[t!]
	    \centering
     \begin{subfigure}{\columnwidth}
        \includegraphics[width=\columnwidth]{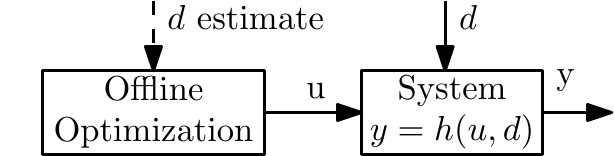}
	    \caption{Block diagram of Offline Optimization.}
	    \label{fig:block_diagram_offline_opt}
     \end{subfigure}
	    \begin{subfigure}{\columnwidth}
	    \includegraphics[width=\columnwidth]{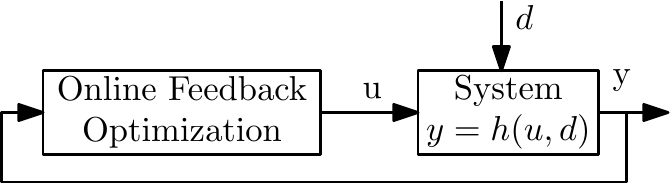}
	    \caption{Block diagram of Online Feedback Optimization.}
	    \label{fig:block_diagram_OFO}    
	    \end{subfigure}
	    \caption{Comparison of Offline Optimization and Online Feedback Optimization.}
        \label{fig:OFO_vs_OPF}
\end{figure}

\ac{OFO} is a method to solve optimization problems using measurements instead of models. This means it is a feedback control method instead of a model-based feedforward approach, compare Figure~\ref{fig:OFO_vs_OPF}. The advantage is that a system model to evaluate $v=h(q,d)$ is not needed. Therefore, no cable and line parameters nor the topology need to be known, and also no active and reactive generation and consumption $d$ need to be measured or estimated. The only model information needed is $\nabla_q h(q,d)$, where $\nabla$ is the gradient operator. It describes how a small change in the reactive power injections $q$ will change the voltage $v$. Note that, this is not the same as knowing which voltage $v$ will result for a specific $q$. We only need to know the derivative of $v$ with respect to $q$. This is very similar to power transfer distribution factors which describe how a change in active power injections will change the line flows. We will from now on refer to this relationship between the effect of a change in $v$ for a small change in $q$ as the sensitivity.

Now, we explain how to drive a power system to the optimal solution of an optimization problem using feedback. To do so we turn an optimization algorithm into a feedback controller, which is the core idea of \ac{OFO}.
This enables us to profit from closed-loop feedback control advantages such as robustness to disturbances $d$ and model mismatch in the sensitivity. This has been done with several different optimization algorithms which all lead to a different system behavior with specific advantages. For an overview see \cite{hauswirth2021optimization}.
Here we explain the idea with an illustrative example, i.e. an optimization problem with no constraints:
\begin{align}\label{eq:op_problem_dummy}
	       \begin{split}
	        	        \min_{q} f(q)
	        \end{split}
	        \end{align}
and the optimization algorithm is gradient descent. This means to minimize a function $f(q)$ one takes gradient steps with step size $\alpha$. The gradient of $f(q)$ is $\nabla f(q)$ and a gradient step with step size $\alpha$ is:
 \begin{align}
	\label{eq:grad_descent}
	\begin{split}
	    q(k+1) &= q(k) - \alpha \nabla f(q(k)).
	\end{split}
	\end{align}
This is an integral controller which keeps changing $q$ until the gradient of the cost function $\nabla f(q(k))$ is zero and therefore $q$ is driven to a locally optimal solution of~\eqref{eq:op_problem_dummy}. Just as with standard integral controllers and due to using feedback this works for a wide range of gains $\alpha$.
An \ac{OFO} controller based on standard gradient descent like~\eqref{eq:grad_descent} does not satisfy any constraints.
To ensure that the constraints on the reactive power injections $q$ and voltages $v$ are satisfied, we can use projected gradient descent. An OFO controller derived from projected gradient descent was presented in \cite{haberle2020non}. We tailor the update law to our specific use case and get
 \begin{equation} \label{eq:verenas_algorithm} 
    q(k+1) = q(k) + \alpha\sigma(q(k),v(k))
\end{equation}
\begin{align}\label{eq:projection_QP}
\begin{split}
    \sigma(q,v) = \arg\min_{w\in\bbR^p} \, &\| w + \nabla f(q)  \|^2
    \\ 
    \text{s. t.} \quad &q_{min} \leq (q+ \alpha w) \leq q_{max} \\ & v_{min} \leq (v+ \alpha \nabla_q h(q,d)w)\leq v_{max}
    \end{split}
\end{align}
with $p$ being the number of reactive power setpoints.
This is also an integral controller and it drives $\sigma(q,v)$ to zero, which is only zero when either $\nabla f(q)$ is zero or if the cost function $f(q)$ cannot be further decreased because constraints on $q$ or $v$ are reached. In both cases, the controller has iteratively changed $q$ until a local optimum has been reached, which is exactly what the controller is designed for. In our implementation in the distribution grid, we were not able to control the reactive power injections directly. However, we can control the power factor $\cos(\phi)$ of the power injections instead. The commands we can send are 0.8 ind., 0.85 ind., 0.9 ind., 0.95 ind., 1, 0.95 cap., 0.9 cap., 0.85 cap., 0.8 cap. This means our control input has to be a discrete value which we can enforce by adding integer constraints. We adapt the controller proposed in~\cite{ortmann2023online} which results in: 
\begin{equation} \label{eq:OFO_with_integer}
    \cos(\phi)(k+1) = \cos(\phi)(k) + \alpha\sigma(\cos(\phi(k)),v(k))
\end{equation}
\begin{align}\label{eq:projection_MIQP}
\begin{split}
    \sigma(\cos&(\phi),v) = \arg\min_{w\in\bbR^p} \, \| w + \nabla f(q)  \|^2
    \\ 
    \text{s.t.} \quad &\cos(\phi)_{min} \leq (\cos(\phi)+ \alpha w) \leq \cos(\phi)_{max} \\
    & v_{min} \leq (v+ \alpha \nabla_{cos(\phi)} q(\cos(\phi), p) \nabla_q h(q,d)w)\leq v_{max}\\
    &\frac{w}{0.05} \in \mathbb{Z},
    \end{split}
\end{align}
where $\mathbb{Z}$ is the set of all integer variables and therefore w/0.05 can only take values of 0.05, -0.05, 0.1, etc. This is the control algorithm we implement on the distribution grid.\\
Problem~\eqref{eq:projection_MIQP} is a \ac{MIQP} that needs to be solved at every time step. Without integer constraints, it is easy and fast to solve even for large systems. With integer constraints, the problem is harder to solve but easier than solving the overall problem~\eqref{eq:op_problem_AEW} including these integer constraints that the hardware setup demands.

\subsection{Necessary Model Information}

The controller used in the implementation needs to know how a change in the $\cos(\phi)$ setpoint is going to affect the voltage. This can be split into two parts. First how a change in $\cos(\phi)$ changes the reactive power injections $q$ ($\nabla_{\cos(\phi)} q(\cos(\phi))$) and second how the reactive power injections affect the voltage ($\nabla_q h(q,d)$). Such sensitivities can either be derived experimentally by changing an input and observing the change in the output or the same can be done using a simulation model of the grid. Furthermore, the sensitivity can be derived mathematically using the admittance matrix of the grid, and the power flow equations~\cite{bolognani2015fast}.

The sensitivity $\nabla_q h(q,d)$ depends on both the topology and line impedances as well as $q$ and $d$. Therefore, the sensitivity changes over time and is generally hard to compute exactly. One has to work with approximations which poses a challenge to any kind of optimization. In such conditions of uncertainty, \ac{OFO} controllers are particularly effective due to their feedback nature. Approximating the sensitivity with a constant matrix has proven to work well~\cite{ortmann2020experimental} and most importantly, even with an approximate sensitivity, the controller will enforce the constraints on both the input ($q$ or $\cos(\phi)$) and output ($v$) in steady-state. Also, temporary constraint violations~\cite{haberle2020non} and the suboptimality are bounded~\cite{colombino2019towards}.
Last but not least, the sensitivity can be learned online from measurements~\cite{picallo2022adaptive} and \ac{OFO} controllers exist that rely on zeroth order optimization algorithms and therefore do not need any sensitivity~\cite{he2022model}.
 
 \subsection{Possible Extensions}
 \ac{OFO} controllers offer great flexibility and possible extensions. We show some that are helpful in power grids.

 \subsubsection{State estimation}
 Instead of feeding the raw voltage measurements into the \ac{OFO} controller one can run the measurements through a state estimation and provide the result of the state estimation to the controller instead. The convergence of this feedback system was proven in~\cite{picallo2020closing}. This also enables to control voltages that are not directly measured and get estimated instead.
 
 \subsubsection{Time-varying constraints}
 The constraints in the control law~\eqref{eq:projection_MIQP} can be different at every time step. This allows to include time-varying constraints of the overarching optimization problem~\eqref{eq:op_problem_AEW}. For example, with certain inverters, one can directly command reactive power injections $q$. Given that an inverter has a current limit the available capacity for reactive power injections would depend on the active power injections which change over time. In other applications, time-varying constraints could be dynamic line ratings or they could be used to temporarily block taps changers, make the controller reduce the power flow on a line, or lower the voltage angle over a circuit breaker.
 
 \subsubsection{Updating the sensitivity}
 The sensitivity depends on the topology, tap changer positions, line parameters, generation, and consumption. These may change over time and therefore also the sensitivity can change over time. If for example, the topology has changed the sensitivity could be recomputed or the results of a new state estimation could be used to update the sensitivity.

\section{Distribution Grid Deployment}

\subsection{Hardware and Communication Setup}
\begin{figure}
    \centering
    \includegraphics[trim= 0 0cm .5cm 0, clip, width=\columnwidth]{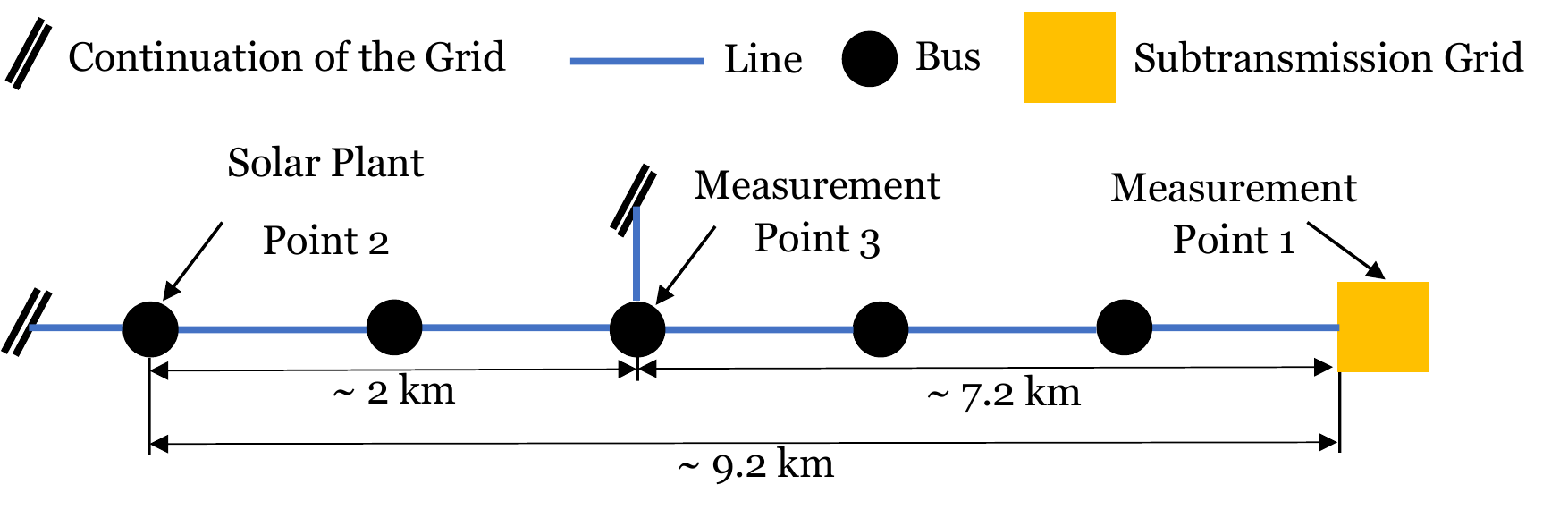}
    \caption{The part of the grid controlled by \ac{OFO} with the connection to the subtransmission grid operator, the measurement points, and the grid topology.}
    \label{fig:grid_AEW}
\end{figure}

\begin{figure}
	\centering
	\includegraphics[width=\columnwidth]{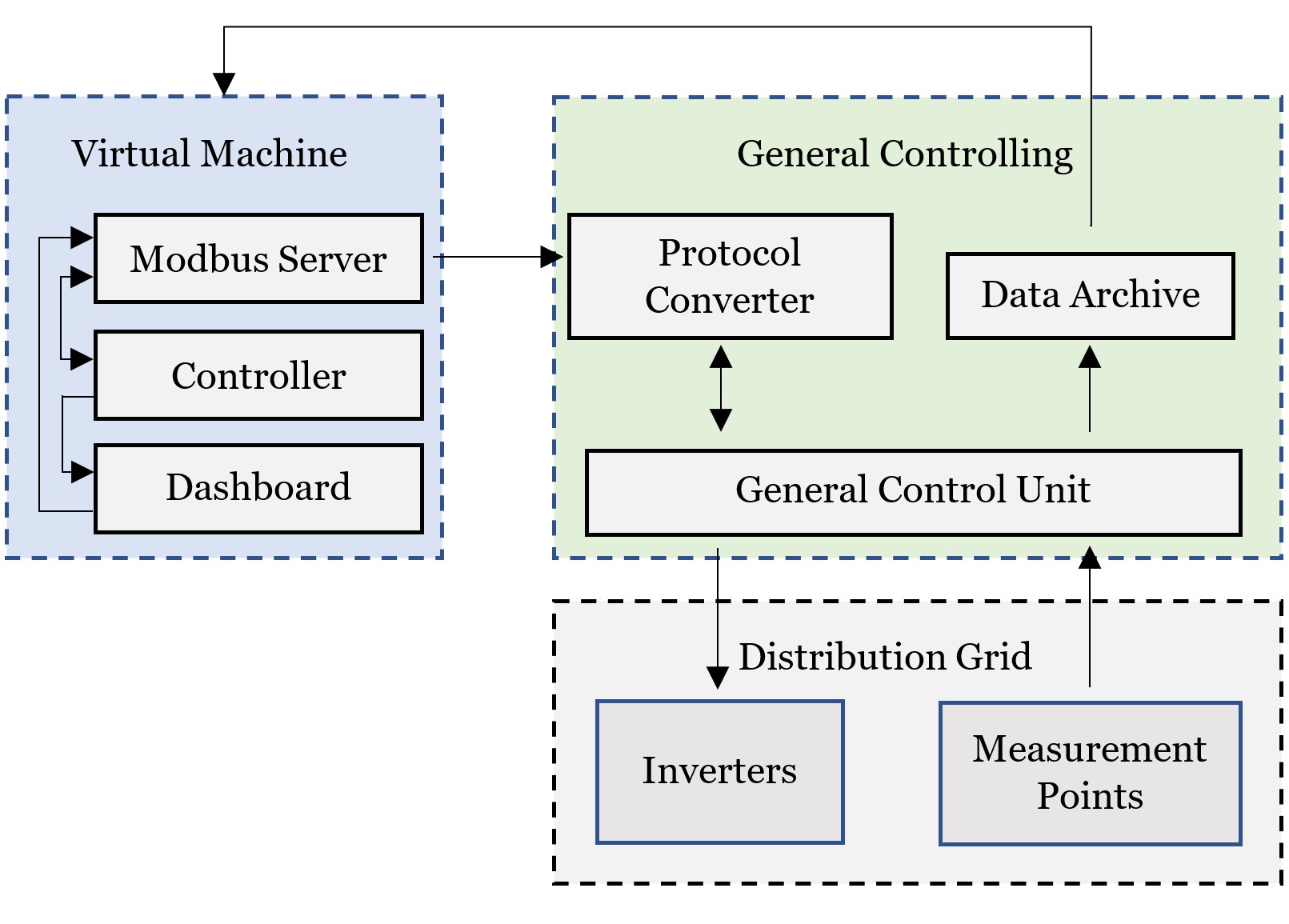}
	\caption{High-level overview of the system including interfaces and communication links.}
	\label{fig:overview_grid}
\end{figure}

\begin{figure}
	\centering
	\includegraphics[width=\columnwidth]{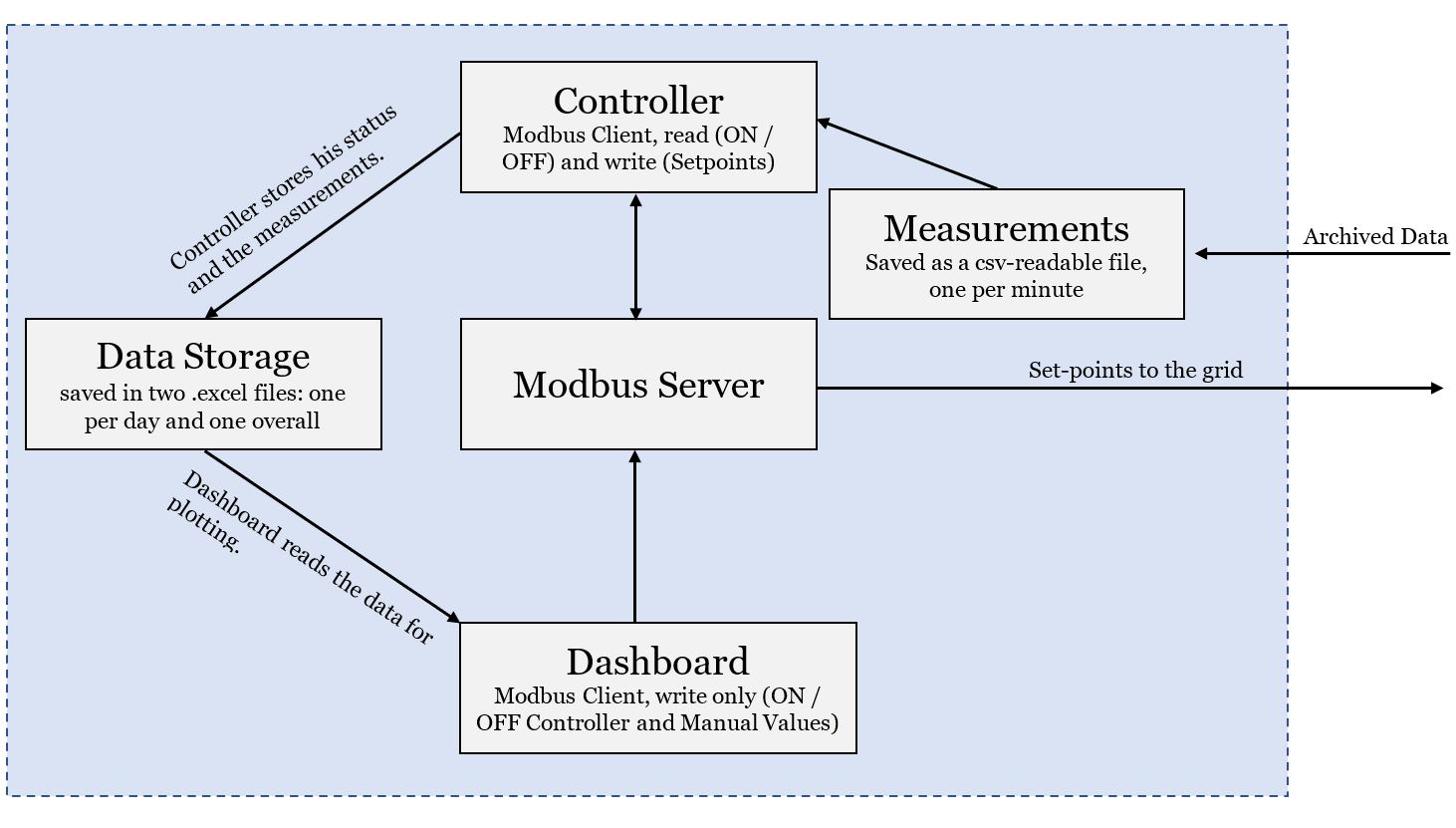}
	\caption{Overview of the programs inside the virtual machine.}
	\label{fig:overview_vm}
\end{figure}

The area of the grid under control by the \ac{OFO} controller is visualized in Figure~\ref{fig:grid_AEW}. We control 16 inverters located at point 2 which is approximately 9.2\,km away from the connection to the subtransmission grid. Their total rated apparent power is 800\,kVA and with our maximum power factor of 0.8, this corresponds to 640\,kW and 480\,kVAr. These inverters operate at 400\,V and are located close to a transformer which transforms the power to the 16\,kV radial grid whose topology is depicted in the figure. Voltage magnitude and power measurements are taken throughout the grid and communicated to the SCADA system of the distribution grid operator.\\
To implement our controller we retrofitted this infrastructure as depicted in Figure~\ref{fig:overview_grid}.  Our controller gets measurements from the SCADA system through an existing Archive Data Server in a CSV file once every minute. It then calculates the new power factor setpoints for the inverters which are collected in a data storage and given to a Modbus server. This Modbus server communicates the setpoints to a protocol converter which transmits the setpoints to the SCADA system over an IEC104 protocol. We equip the inverters with a Siemens Smartgrid-Box A8000 to be able to send them these setpoints. The SCADA system communicates with this A8000 through an IEC~60870-5-104 protocol. Data logging at the inverters is done with an ADL-MXSpro from Meier-NT. A dashboard visualizes the measurements and setpoints and it can be used to enable, disable, and reset the controller or for manually choosing the setpoints. To enable these features the dashboard crawls data from the data storage and communicates with the Modbus server.\\
The \ac{OFO} controller, the dashboard, and the data storage are implemented on a virtual machine inside the control room. Figure~\ref{fig:overview_vm} shows an overview of the programs running and interacting on the virtual machine. The code was written in Python and its execution is computationally very light, meaning no large computation power is needed.

\subsection{Controller Setup}
\begin{figure}
    \centering
    \includegraphics[width=\columnwidth]{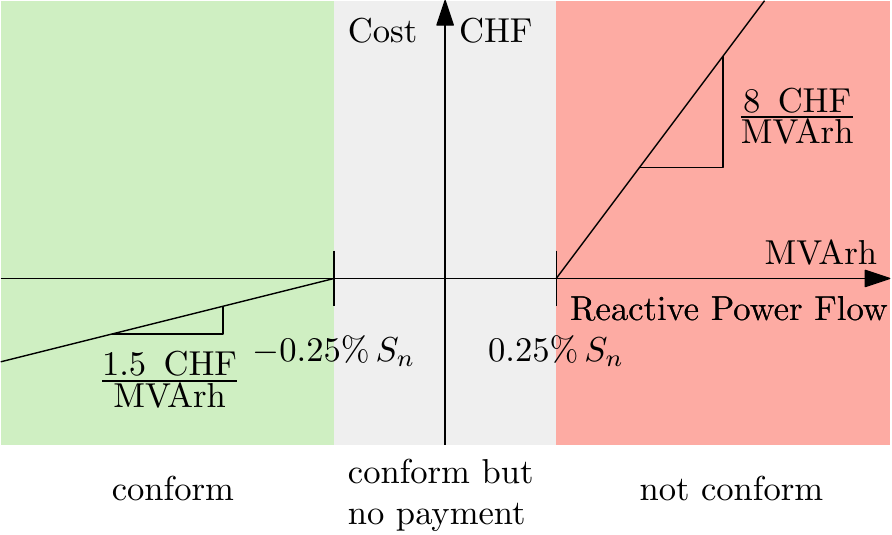}
    \caption{Cost function for reactive power flows into the subtransmission grid based on the pricing scheme of the subtransmision grid operator. The distribution grid operator has to pay a high penalty for capacitive flows and gets a small amount of money for inductive flows.}
    \label{fig:cost_function_AEW}
\end{figure}

The controller is implemented as follows. The SCADA system provides the controller with voltage magnitude measurements of the three points shown in Figure~\ref{fig:grid_AEW}. At measurement point~1 we also get the reactive power flow which is needed to calculate $\nabla f(q)$. The goal is to optimize the reactive power flow at the connection point to the external grid (measurement point~1).
The cost function $f(q)$ is based on the pricing scheme of the subtransmission grid operator and it is a piece-wise linear function, see Figure~\ref{fig:cost_function_AEW}. Due to the linearity the derivative $\nabla f(q)$ is constant in each area. There is a high cost for capacitive reactive energy and a small reward for inductive reactive energy (MVArh). A deadband with no cost nor reward exists and is of size 0.25\%~$S_n$ where $S_n$ is the sum of the apparent power of all transformers at the connection point to the subtransmission grid. Recall that, $\sigma = \nabla f(q)$ when no constraints are active and note that the derivative of the cost function $\nabla f(q)$ is zero in the gray area. Hence, $\sigma$ would therefore be zero in the gray area (as long as there are no voltage violations) and the controller would not change the setpoints. To circumvent this, we artificially change the cost function to have a small gradient in the gray area which ensures that the controller tries to drive the reactive power flows into the conform (green) area.
The sensitivity $\nabla_q h(q,d)$ was calculated based on a model and is kept constant.

Given that our control approach is relying on communication infrastructure, it is necessary to define a fallback strategy in case the communication breaks down. In case the controller does not receive measurements for five minutes it instructs the inverter to operate at a power factor of 1. If the inverters do not receive commands from the controller anymore, they also set their power factor to 1.

\section{Results and Data Analysis}
In this section, we analyze the consequences of using off-the-shelf hardware, the interaction of an \ac{OFO} controller with a real distribution grid, and the behavior of the controller. The controller went live in December~2022 and the data analysis of the first months revealed the following.

\begin{figure}
	\centering
	\setlength\fwidth{\columnwidth}
	\setlength\fheight{6cm}
	\begin{adjustbox}{max width=\columnwidth}
		\input{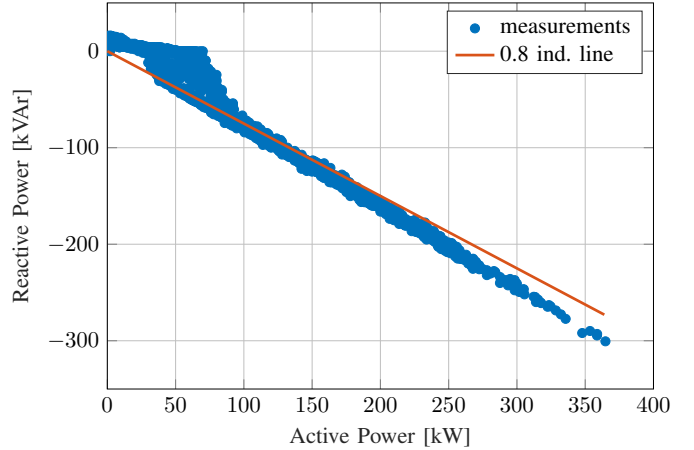}
	\end{adjustbox}
	\caption{Active and reactive power injections of the inverters in the month of January together with a line indicating a power factor of 0.8 inductive.}
	\label{fig:active_vs_reactive_AEW}
\end{figure}

The controller gives power factor setpoints $\cos(\phi)$ to the inverters. Figure~\ref{fig:active_vs_reactive_AEW} shows the active and reactive power injections of the inverters for a power factor setpoint of 0.8 inductive. The figure shows that the inverters do not track the setpoint well. Especially for low active power injections the reactive power injection is capacitive even though the controller was asking for a power factor of 0.8 inductive. This happens because old norms only required reactive power tracking for active power generation of larger than 5\% of the rated power. These measurements highlight how important it is to utilize measurements as feedback because not only a grid model can be wrong, but also actuators might not follow their reference.

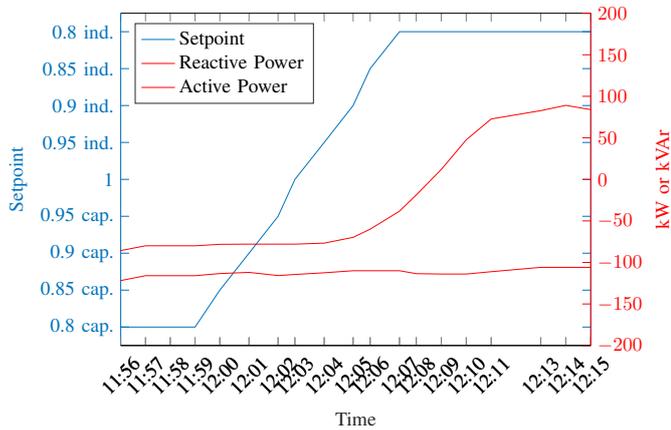
\begin{figure}
	\centering
	\setlength\fwidth{\columnwidth}
	\setlength\fheight{6cm}
	\begin{adjustbox}{max width=\columnwidth}
		\input{Figures/cost_function_change}
	\end{adjustbox}
	\caption{Change of the setpoint and reactive power when the cost function changes at noon. Data point at 12:12 is missing.}
	\label{fig:cost_function_change}
\end{figure}

In 2022, the pricing scheme of the subtransmission grid operator was different from the one in Figure~\ref{fig:cost_function_AEW} and was dependent on the time of day. Therefore, the optimization problem changed at certain times and the controller automatically adjusted the setpoints to drive them to the optimal solution of the new optimization problem. Figure~\ref{fig:cost_function_change} shows the behavior of the controller when the cost function changes. Note that, the controller iteratively changes the setpoints over several steps. This iterative behavior is at the core of \ac{OFO} as it allows to work with minimal model information. It can also be seen that there is a time delay of approximately four minutes between the setpoints being changed and the inverters adjusting their reactive power injections. The presence of this time delay means that voltage violations could persist for up to four minutes before the controller is able to mitigate them. Currently, the VDE~4105 norm is allowing temporary voltage violations for up to one minute~\cite{VDE}. We conclude that a sampling time of the controller of fewer than 10~seconds should be chosen for future implementations to guarantee that voltage violations are cleared within one minute.

The sensors in the grid only send a new measurement to the control room when the measured value has changed by more than 1\%. The gathered data suggests that the controller does not have a problem with the measurements being triggered.

Note that, the distribution grid is currently not experiencing voltage violation and hence the virtual reinforcement feature of the controller could not be evaluated.

\section{Conclusion}
In this paper, we presented the retrofitting of existing grid infrastructure to optimize reactive power flows in a distribution grid using an \ac{OFO} controller that controls reactive power injections of PV inverters. The controller also enforces voltage limits by adjusting the reactive power injections which means there exists potential to virtually reinforce the grid by mitigating voltage limit violations. The implementation shows that the controller is robust against model mismatch, is compatible with the legacy grid infrastructure, and can work with triggered measurements. We consider this 24/7 implementation to be a system prototype demonstration in an operational environment and conclude that the \ac{OFO} control method has therefore reached technology readiness level 7.\\
Further investigations are needed to quantify the monetary value of the virtual grid reinforcement that voltage control through reactive power can provide. Also, given the high technology readiness level, \ac{OFO} might be considered for commercial control room software. Finally, the principle of defining a control problem as an optimization problem and then using an \ac{OFO} controller to solve the optimization and therefore the control problem could be applied to more problems, e.g. active power curtailment, curative actions and automatic redispatch, disaggregation of flexibility commands onto several resources.

\bibliographystyle{IEEEtran}
\bibliography{IEEEabrv,biblio}

\end{document}

%% file: Figures/cost_function_change.tex
\definecolor{mycolor1}{rgb}{0.00000,0.44700,0.74100}%
\begin{tikzpicture}

\begin{axis}[%
width=0.951\fwidth,
height=\fheight,
at={(0\fwidth,0\fheight)},
scale only axis,
xmin=1.00069734551653,
xmax=1.01392810900143,
xtick={1,1.00069734551653,1.00139451065479,1.00209203718987,1.00278974476532,1.0034865479829,1.00430795534339,1.00512628727301,1.00560064955789,1.00642476835492,1.00723966478836,1.00771890986653,1.00854483777221,1.00901630552107,1.00972034270671,1.01042311401397,1.01112516194553,1.01252148114872,1.01322533770144,1.01392810900143,1.01462979523785},
xticklabels={{11:55},{11:56},{11:57},{11:58},{11:59},{12:00},{12:01},{12:02},{12:03},{12:04},{12:05},{12:06},{12:07},{12:08},{12:09},{12:10},{12:11},{12:13},{12:14},{12:15},{12:16}},
xticklabel style={rotate=45},
xlabel style={font=\color{white!15!black}},
xlabel={Time},
separate axis lines,
every outer y axis line/.append style={mycolor1},
every y tick label/.append style={font=\color{mycolor1}},
every y tick/.append style={mycolor1},
ymin=-4.5,
ymax=4.5,
ytick={-4,-3,-2,-1,0,1,2,3,4},
yticklabels={{0.8 cap.},{0.85 cap.},{0.9 cap.},{0.95 cap.},{1},{0.95 ind.},{0.9 ind.},{0.85 ind.},{0.8 ind.}},
ylabel style={font=\color{mycolor1}},
ylabel={Setpoint},
axis background/.style={fill=white}
]
\addplot [color=mycolor1, forget plot]
  table[row sep=crcr]{%
1	-4\\
1.00069734551653	-4\\
1.00139451065479	-4\\
1.00209203718987	-4\\
1.00278974476532	-4\\
1.0034865479829	-3\\
1.00430795534339	-2\\
1.00512628727301	-1\\
1.00560064955789	0\\
1.00642476835492	1\\
1.00723966478836	2\\
1.00771890986653	3\\
1.00854483777221	4\\
1.00901630552107	4\\
1.00972034270671	4\\
1.01042311401397	4\\
1.01112516194553	4\\
1.01252148114872	4\\
1.01322533770144	4\\
1.01392810900143	4\\
1.01462979523785	4\\
};
\label{plotyyref:leg1}

\end{axis}

\begin{axis}[%
width=0.951\fwidth,
height=\fheight,
at={(0\fwidth,0\fheight)},
scale only axis,
xmin=1.00069734551653,
xmax=1.01392810900143,
xtick={1,1.00069734551653,1.00139451065479,1.00209203718987,1.00278974476532,1.0034865479829,1.00430795534339,1.00512628727301,1.00560064955789,1.00642476835492,1.00723966478836,1.00771890986653,1.00854483777221,1.00901630552107,1.00972034270671,1.01042311401397,1.01112516194553,1.01252148114872,1.01322533770144,1.01392810900143,1.01462979523785},
xticklabels={{11:55},{11:56},{11:57},{11:58},{11:59},{12:00},{12:01},{12:02},{12:03},{12:04},{12:05},{12:06},{12:07},{12:08},{12:09},{12:10},{12:11},{12:13},{12:14},{12:15},{12:16}},
xticklabel style={rotate=45},
every outer y axis line/.append style={red},
every y tick label/.append style={font=\color{red}},
every y tick/.append style={red},
ymin=-200,
ymax=200,
ytick={-200, -150, -100,  -50,    0,   50,  100,  150,  200},
ylabel style={font=\color{red}},
ylabel={kW or kVAr},
axis x line*=bottom,
axis y line*=right,
legend style={at={(0.03,0.97)}, anchor=north west, legend cell align=left, align=left, draw=white!15!black}
]
\addlegendimage{/pgfplots/refstyle=plotyyref:leg1}
\addlegendentry{Setpoint}
\addplot [color=red]
  table[row sep=crcr]{%
1	-85.925529\\
1.00069734551653	-85.726509\\
1.00139451065479	-79.954872\\
1.00209203718987	-79.954872\\
1.00278974476532	-79.954872\\
1.0034865479829	-78.196846\\
1.00430795534339	-77.964653\\
1.00512628727301	-77.964653\\
1.00560064955789	-77.964653\\
1.00642476835492	-76.770523\\
1.00723966478836	-69.778793\\
1.00771890986653	-59.966152\\
1.00854483777221	-38.238159\\
1.00901630552107	-18.734015\\
1.00972034270671	11.941312\\
1.01042311401397	47.610928\\
1.01112516194553	72.712204\\
1.01252148114872	82.746948\\
1.01322533770144	89.083916\\
1.01392810900143	83.93531\\
1.01462979523785	83.93531\\
};
\addlegendentry{Reactive Power}

\addplot [color=red]
  table[row sep=crcr]{%
1	-125.582794\\
1.00069734551653	-121.823013\\
1.00139451065479	-115.951874\\
1.00209203718987	-115.951874\\
1.00278974476532	-115.951874\\
1.0034865479829	-113.364586\\
1.00430795534339	-111.971428\\
1.00512628727301	-115.819191\\
1.00560064955789	-114.492386\\
1.00642476835492	-112.435814\\
1.00723966478836	-109.981216\\
1.00771890986653	-109.981216\\
1.00854483777221	-109.981216\\
1.00901630552107	-113.497276\\
1.00972034270671	-113.961655\\
1.01042311401397	-113.961655\\
1.01112516194553	-111.145058\\
1.01252148114872	-105.914238\\
1.01322533770144	-105.914238\\
1.01392810900143	-105.914238\\
1.01462979523785	-105.914238\\
};
\addlegendentry{Active Power}

\end{axis}
\end{tikzpicture}%